\documentclass[12pt]{article}
\usepackage{amsfonts}
\usepackage{amssymb}
\makeatletter
\renewcommand{\section}{\@startsection{section}{1}{0pt}
  {1.5ex}{1.5ex}{\Large\bfseries}}
\renewcommand{\subsection}{\@startsection{subsection}{2}{0pt}
  {1ex}{1ex}{\large\bfseries}}
\makeatother

\def\bfr{\begin{flushright}}
\def\efr{\end{flushright}}

  \def\bsh{\backslash}

 \newfont{\bbbold}{msbm12}
 \def\bi{\begin{itemize}}
\def\ei{\end{itemize}}
\def\com{\mbox{\bbbold C}}

 \def\cK{{\cal K}}
 
 \def\cM{{\cal M}}
 
 \def\cO{{\cal O}}

 \def\cR{{\cal R}}

 \newfont{\goth}{eufm10 scaled \magstep1}

 \def\gl{\mbox{\goth l}}

 \def\gp{\mbox{\goth p}}

 \def\gs{\mbox{\goth s}}

 \def\a{\alpha}
 \def\b{\beta}
 
 \def\d{\delta}\def\D{\Delta}
 
 \def\f{\phi}\def\vf{\varphi}
 \def\h{\eta}
 \def\k{\kappa}
 \def\l{\lambda}
 \def\m{\mu}
 \def\n{\nu}
 \def\p{\pi}
 
 \def\r{\rho}

 \def\th{\theta}

 \def\adt{\alpha'}
 \def\bdt{\beta'}

 \def\be{\begin{equation}}\def\ee{\end{equation}}
 \def\bea{\begin{eqnarray}}\def\eea{\end{eqnarray}}
 \def\ba{\begin{array}}\def\ea{\end{array}}

 \def\del{\partial}

 \def\str{\rm str}
 \def\xz{\times}

 \def\del{\partial}
 
 \def\bt{\bullet}
 \def\3dt{\dot{3}}

 \let\la=\label

 {}

 \def\nn{\nonumber}
 \def\bd{\begin{document}}
 \def\ed{\end{document}}
 \def\bea{\begin{eqnarray}}\def\barr{\begin{array}}\def\earr{\end{array}}
 \def\eea{\end{eqnarray}}
 \def\ft#1#2{{\textstyle{{\scriptstyle #1}\over {\scriptstyle #2}}}}
 \def\fft#1#2{{#1 \over #2}}
 \newcommand{\eq}[1]{(\ref{#1})}
 \def\eqs#1#2{(\ref{#1}-\ref{#2})}
 \def\det{{\rm det\,}}
 \def\tr{{\rm tr}}\def\Tr{{\rm Tr}}
  \def\str{{\rm str}} \def\diag{{\rm diag}}
 \def\sdet{{\rm sdet}}
\newcommand{\hoch}[1]{$^{#1}$}

\def\cxz{\circ \hspace{-13pt} \diagup \hspace{-13pt}  
\diagdown}
\def\btx{\bt \hspace{-13pt} \diagup \hspace{-13pt}  \diagdown}
\def\fc{\framebox}
\def\rc{}
\def\vs{\vspace{1cm}}
\def\bc{\begin{center}}
\def\ec{\end{center}}
\def\red{\textcolor{myblue1}}
\def\bta{\begin{tabular}}
\def\et{\end{tabular}}

\def\gap{\vspace{1cm}}

\begin{document}

\thispagestyle{empty}

  \hfill{\today}

 \vspace{20pt}

 \begin{center}
 {\Large{\bf Superconformal field theories in analytic superspace }}
 \vspace{30pt}

 {P.J. Heslop\footnote{A substantial part of the work presented here
     was done in collaboration with P.S.Howe to whom I am also very
     grateful for making 
 helpful  comments on this manuscript.}}\\[20pt]

 {\sl \begin{tabular}{ll} &II. Institut f\"ur Theoretische Physik
der Universit\"at Hamburg 
 \\ &(now at Institut f\"ur Theoretische Physik
der Universit\"at Leipzig)\\ \\
\et}

 \vspace{60pt}

 \end{center}

 {\bf Abstract}

We summarise recent work on superconformal field theories using
analytic superspace. All operators of $N=4$ SYM can be
given as unconstrained superfields on analytic superspace. We show how
to write down operators as superfields on  analytic superspace and how
to completely solve the 
Ward indentities for their correlation functions. We discuss
the non-renormalisation of certain operators, and of some of their correlation
functions. We discuss the relationship between harmonic and analytic
superspace. Finally we discuss applications of these techniques to
superconformal field theory in 6 dimensions.

 {\vfill\leftline{}\vfill \vskip  10pt

 \baselineskip=15pt \pagebreak \setcounter{page}{1}

\section{Introduction}

Superconformal field theories have acquired considerable interest in
recent years due to the discovery of the AdS/CFT correspondence
which relates them to supergravity or superstring
theories on anti de-Sitter (AdS) spaces~\cite{Maldacena:1998re}.
In particular $N=4$ super Yang-Mills (SYM) provides the clearest and
most concrete example of the AdS/CFT 
correspondence, via which it is related to IIB supergravity or string theory
on an $AdS_5\xz S^5$ background. $N=4$ SYM is also of fundamental
interest in its on right due to a number of remarkable properties: it
has the largest possible amount of flat space supersymmetry, it is
uniquely determined by the coupling constant and the gauge group and
it is superconformally invariant even as a quantum theory~\cite{Sohnius:sn}. It is thus
the most symmetric known four dimensional gauge theory, thus providing
an important testing ground for more physicaly interesting gauge
theories. In particular, one would
like to know to what extent the theory is determined by
its symmetries. 

The $(2,0)$ tensor multiplet in six dimensions is also of great
interest as it is dual to $M$-theory on $AdS_7\xz S^4$ via the AdS/CFT
correspndence. It is much more mysterious than $N=4$ SYM, however, as
the classical theory is not known and there is no dimensional coupling
constant. Nevertheless one can still assume a conformally invariant
quantum theory and investigate the consequences of conformal invariance.

Harmonic/analytic superspaces provide the clearest way 
to answering these questions since the full  superconformal symmetry
is manifest, and it acts on analytic superfields. In particular it
is very easy to write down conformally invariant correlation functions.

In this talk we give an introduction to harmonic/ analytic
superspace and its applications in superconformal quantum field
theory. 
 We will begin in section~\ref{sec:Has} with the standard harmonic superspace techniques
 which allow one to write certain supermultiplets as analytic superfields
 (without superindices).
We then discuss some  technical details of supercoset spaces and super
  Dynkin diagrams. We show how one can
 obtain all  superconformal fields as {{ unconstrained}} analytic
 superfields. This is achieved with the aid of 
superindices (ie non-trivial linear representations of supergroups).
The techniques are very general: any unitary irreducible
superconformal 
  representation can be given as an analytic superfield on any (non-twistor)
  superspace. 
In section~\ref{sec:SYM} we consider some applications of this technique in  $N=4$
SYM.
Section~\ref{sec:hss}} deals with the relationship between harmonic
superspace and analytic superspace, showing how to lift unconstrained 
fields on analytic superspace to constrained fields on harmonic
superspace.
Finally, in section~\ref{6d}, we look very briefly at some
applications to six-dimensional  superconformal field theory.


\section{Harmonic/ Analytic superspace}\la{sec:Has}

Harmonic superspace $\cM_H$ is obtained by appending an internal manifold
$K$ to
extended Minkowski superspace $\cM$ 
\be \cM_H=\cM \xz K.
\ee
The internal manifold is usually a coset manifold of the internal
group (for example $SU(N)$ for N-extended supersymmetry in 4
dimensions) and we write the coordinates of $K$ accordingly as
$u_I{}^j$ where
$u_I{}^j$ are matrix elements of the internal group. So the
coordinates of harmonic superspace are $(x,\th^i,\bar \th_i,u_I{}^j)$.
%
%
Harmonic superspace was first introduced by GIKOS in
1984~\cite{Galperin:1984av} for the case of $N=2$ supersymmetry.
In this case the internal group is $SU(2)$ carried by the
indices $i=1,2$. The internal manifold is given by the coset $ U(1)
\bsh SU(2)$ which is just the Riemann sphere and can also be
described by the coset $P\bsh SL(2;\com)$ where $P$ is the set of 2x2
complex lower triangular matrices with unit determinant.
%
%
A coset representative for this coset is given by
\be
s(y)=\left(
  \ba{cc} 1 &y\\0 &1 \ea \right) 
\ee
where $y$ is a complex number which covers all but one point of the
sphere $K$. In order to cover the whole space one has to consider
the new coordinate $y'=1/y$ in the standard way for the Riemann sphere.

There are two alternative but equivalent ways of dealing with the 
internal coordinates $u_I{}^j$. One can 
consider them to have any values in the entire group manifold
$SU(2)$ (or $SL(2;\com)$). By restricting oneself to superfields with special covariance
properties this is equivalent to considering fields on the
coset. Alternatively one can consider fields on harmonic
superspace directly in which case one considers the internal
coordinates $u_I{}^j=s(y)_I{}^j$. One must then also consider the
other coordinate patch with coordinates $y'$.
We adopt the latter point of view since it allows easier
generalisation to the case of analytic superfields which transform
non-trivially under supergroups (see section~\ref{transf}).

\subsection{Why is harmonic superspace useful?}\la{Y}

We will illustrate the usefulness of harmonic superspace through an
example, the hypermultiplet in $N=2$ supersymmetry. The hypermultiplet 
consists of 4 real scalars $f_i(x)$, and two spinors
$\psi_{\a}(x), \k^{\adt}(x)$. These fields can be packaged into a single superfield 
$q_i$ on $N=2$ Minkowski superspace satisfying the constraint
\be
D_{\a (i}q_{j)}=\bar D_{\adt (i} q_{j)}=0
\quad \Rightarrow \quad 
q_i(x,\th,\bar \th)= f_i(x) + \th_i^{\a} \psi_{\a}(x) + \bar
\th^{\adt}_i \bar \k_{\adt}(x) + \dots \la{con}
\ee
where the dots indicate further terms which involve no further fields, and
all component fields satisfy there equations of  motion. 
On $N=2$ harmonic superspace,  the superfield $q_i$ becomes a harmonic 
superfield, $q(x,\th,\bar \th,u):=u_1{}^iq_i= q_1 +yq_2$ and the constraints~\eq{con} become simple Grassmann
analyticity conditions
$D_{\a 1} q
  = \bar D^2_{\adt} q=0$
where $D_{\a 1}:=u_1{}^i D_{\a i},\ \bar D^2_{\adt}:=\bar
  D_{\adt}^i (u^{-1})_i{}^2.$ 
Note that we have imposed analyticity on the
internal manifold, H-analyticity, which leads to the short expansion in
the internal coordinates. The Grassmann analyticity conditions are the generalisation of
chirality constraints in $N=1$ and as in that case one
can easily solve the constraints.  The solution here is
$q=q(x_A,\th^2,\bar \th_1,y)$
with a suitably redefined $x$ coordinate $x_A$ and with
$\th^2=\th^i(u^{-1})_i{}^2,\ \bar \th_1=u_1{}^i\bar \th_i$. 
The hypermultiplet  can therefore be thought of as an
analytic superfield on analytic superspace which has 
  coordinates $(x_A,\th^2,\bar \th_1,y)$. 

Strictly speaking analytic superspace is only defined when one
  complexifies the coordinates, so that $x$ is now complex and $\th$ is
  unrelated to $\bar \th$ (y is already complex). To indicate this we
  will denote $\bar \th$ by $\vf$ from now on. We can always go back
  to the real case at the end of any calculation simply by considering
  $x$ real and $\vf=\bar \th$.

\subsection{Supercosets and super Dynkin diagrams}

A further advantage of considering the complex case is that all
superspaces are then supercosets of the (complex) superconformal
group $SL(4|N;\com)$ allowing us to use many techniques from
representation theory and parabolic induction.
We use a non-standard representation for a matrix in
the Lie algebra $\gs \gl(4|N)$, corresponding to changing the basis
on which the matrix acts from $(4|N)$ to $(2|N|2)$.
In this basis all superspaces will have the form $P\bsh SL(4|N)$ with
$P$ a
block lower triangular matrix~\cite{Heslop:2001zm}.

The super Dynkin diagram of the superconformal group corresponding to
this choice of basis is 
\be
\begin{picture}(230,0)(0,-10)
\put(0,0){\makebox[0pt][l]{$\bt\hspace{1.5em}\circ\hspace{1.5em}
    \underbrace{\bt \hspace{1.5em}
\bt\hspace{1.5em}\cdots \hspace{1.5em}\bt\hspace{1.5em}\bt}_{N-1}\hspace{1.5em}\circ\hspace{1.5em}
\bt$} 
\rule[.5ex]{6.45em}{.1ex} 
$\hspace{4.5em}$
\rule[.5ex]{6.4em}{.1ex}
 } 
\end {picture}
\ee
Here the $N-1$ central black nodes
represent the internal $\gs \gl(N)$ subalgebra ($su(N)$ in the real case),
the two black nodes on the ends
represent
the space-time $\gs \gl(2)$ Lorentz group representations. 
The two white nodes
represent odd roots in the Lie superalgebra.

We now consider coset spaces of the superconformal group $P\bsh
SL(4|N)$. In the case where $P$ is a parabolic subgroup (corresponding
to a block lower triangular matrix) these can be represent by
putting crosses on the Dynkin diagram~\cite{Baston}. 

In fact all superspaces associated with four-dimesnional Minkowski space can be
represented in this way~\cite{Howe:1995md}.

Irreducible representations transform under the Levi subgroup $L \subset
P$ which corresponds to the
`block diagonal part' of the block lower triangular matrices of $P$.

\subsection{Examples of $N=2$ superspaces $P\bsh SL(4|2)$}

We illustrate the above points using a table containing some examples
in the $N=2$ case.

\begin{tabular}{|c|c|c|c|}\hline \rule[-3mm]{0mm}{8mm}
{superspace}&$P, L$& {coordinates} & {Dynkin diagram}
\\\hline
{$\ba{c} \rm{Minkowski}\\{(x,\th^i,\vf_i)}\ea$}& {\tiny $\left( \ba{cc|cc|cc} \bt &\bt&&&&\\\bt &\bt&&&&\\\hline
\circ &\circ&\bt&\bt&&\\\circ &\circ&\bt&\bt&&\\\hline \circ &\circ&\circ&\circ&\bt&\bt\\\circ &\circ&\circ&\circ&\bt&\bt \ea \right)$}&
{\small $\left( \ba{c|c|c} 1^{\a}{}_{\b}& {\th^{\a i}}&{x^{\a \bdt}} \\\hline
& 1_i{}^j& {\vf_{i}{}^{\bdt}} \\\hline
&& 1_{\adt}{}{\bdt} \ea \right)$}&
$\ba{c} \begin{picture}(70,10)
\put(-10,0){\makebox[0pt][l]{\small $\bt\hspace{1em}\otimes\hspace{1em}\bt
    \hspace{1em}
\otimes \hspace{1em}\bt$}} 
\put(-7,0){\rule[.5ex]{80pt}{.1ex}}
\end {picture}\\
\mbox{\small $\gl=sl(2)^3 \oplus C \hspace{-9pt} C^2$} \ea$
\\[25pt]
{$\ba{c} \rm{Harmonic}\\{(x,\th^i,\vf_i,y)}\ea$}&{\tiny $\left( \ba{cc|cc|cc} \bt &\bt&&&&\\\bt &\bt&&&&\\\hline
\circ &\circ&\bt&&&\\\circ &\circ&\circ&\bt&&\\\hline \circ &\circ&\circ&\circ&\bt&\bt\\\circ
&\circ&\circ&\circ&\bt&\bt \ea \right)$}&
{\small $\left( \ba{c|cc|c} 1^{\a}{}_{\b}& {\th^{\a 1}}& {\th^{\a 2}}&{x^{\a \bdt}} \\\hline
& 1& {y}& {\vf_{1}{}^{\bdt}} \\
&  &1 &{\vf_{2}{}^{\bdt}} \\\hline
&&& 1_{\adt}{}{\bdt} \ea \right)$}&
$\ba{c} \begin{picture}(70,10)
\put(-10,0){\makebox[0pt][l]{\small $\bt\hspace{1em}\otimes\hspace{1em}\times
    \hspace{1em}
\otimes \hspace{1em}\bt$}} 
\put(-7,0){\rule[.5ex]{85pt}{.1ex}}
\end {picture}\\ 
\mbox{\small $\gl=sl(2)^2 \oplus C \hspace{-9pt} C^3$}
\ea$
\\[30pt]
{$\ba{c} \rm{Analytic}\\{(x,\th^2,\vf_1,y)}\ea$}&{\tiny $\left( \ba{cc|cc|cc} \bt &\bt&\bt&&&\\\bt &\bt&\bt&&&\\\hline
\bt &\bt&\bt&&&\\\circ &\circ&\circ&\bt&\bt&\bt\\\hline \circ &\circ&\circ&\bt&\bt&\bt\\\circ
&\circ&\circ&\bt&\bt&\bt \ea \right)$}&
{\small $\left( \ba{c|cc|c} 1^{\a}{}_{\b}&& {\th^{\a 2}}&{x^{\a \bdt}} \\\hline
& 1& {y}& {\vf_{1}{}^{\bdt}} \\
&  &1 & \\\hline
&&& 1_{\adt}{}^{\bdt} \ea \right)$}&
$\ba{c} \begin{picture}(70,10)
\put(-10,0){\makebox[0pt][l]{\small $\bt\hspace{1em}\ominus\hspace{1em}\times
    \hspace{1em}
\ominus \hspace{1em}\bt$}} 
\put(-7,0){\rule[.5ex]{85pt}{.1ex}}
\end {picture} \\ 
\mbox{\small $\gl=sl(2|1)^2 \oplus C \hspace{-9pt} C $}
\ea$\\
\hline
\et

This table shows various aspects of the supercoset representation of
the three $N=2$ superspaces, Minkowski, harmonic and analytic
superspace.
In particular each space has the form $P\bsh SL(4|2)$ and the second
column of this table gives the subgroup $P$: $P$
consists of matrices which have non-zero entries where there is a
black or white circle and all other 
entries are zero. $P$ is always block lower triangular. Fields on this
space will transform under the Levi 
subgroup $L$ which is the set of matrices which have non-zero
entries only where there are black nodes. This is the block diagonal
part of $P$. The coordinates of the space can be thought of as lying
in the zero components of $P$ as indicated in the third column. Finally 
one can give a corresponding Dynkin diagram with crosses through. The
Dynkin diagram gives another simple way to read off the Levi subalgebra $\gl$ corresponding to $L$: crossed
though nodes give $\com$ charges and the remaining Dynkin diagram
after these nodes are taken away gives the rest of $\gl$. 

In particular note that for analytic superspace $\gl$ is
itself a superspace~\cite{Howe:1995md,Lukierski:1988vw}. This is indicated by the
Dynkin diagram: on removing the crossed through node (corresponding to 
the $\com$ charge) one is left with two disconnected Dynkin diagrams
each representing the superalgebra $\gs \gl(2|1)$.




\subsection{Superfields on harmonic / analytic superspace}

Superfields on the above spaces carry representations of the
superconformal group $SL(4|N)$. These representations can be specified
by putting
{Dynkin labels} - specifying the highest weight state - above the
Dynkin nodes. 

Representations of the N-extended superconformal group are usually specified
by the following labels
\be (\D,R,J_1,J_2;a_1 \dots a_{N-1})\la{qn}\ee
where $\D$ is the dilation weight, $J_1,J_2$ the Lorentz spin, $R$ the 
$R$ charge and $a_i$ are Dynkin labels specifying the representation
of the internal group $SL(N)$. These numbers are related in a
straightforward manner to the $N+3$ super
Dynkin labels $n_i$ which one puts above the nodes of the Dynkin
diagram: the two extremal nodes are related to the Lorentz spin
$n_1 = 2J_1$,  $n_{N+3}=2J_2$ the two odd nodes are given by the
linear combination
             $n_2 = {1\over2}(\D-R) + J_1 + {m\over N} -m_1$,
             $n_{N+2} = {1\over2}(\D+R)+J_2-{m\over N}$ and the
             central nodes are given by the Dynkin labels of the
             internal group $SL(N)$. 
            Here $ m:=\sum_{k=1}^{N-1}k a_k$ and
$m_1:=\sum_{k=1}^{N-1}a_k \label{m}$.
We can now describe {any
unitary irreducible representation} as a superfield on {any
superspace} and also its transformation properties under the
superconformal group.

Given a representation (specified for example by giving its quantum
numbers as in~\eq{qn}) and a superspace, one must 
work out the Dynkin labels (as shown above) and write them above the
Dynkin diagram and
put crosses on the Dynkin diagram as dictated by the superspace in
question. 
The resulting Dynkin diagram then tells you how the field transforms
  on this superspace. Technically, one has to convert the Dynkin
  diagrams 
  to Young tableaux using a simple formula and use (super)indices
  which are symmetrised as
  dictated by the Young tableau.  

However, since our spaces are
non-compact, the superfields one produces in this way are often not
irreducible:
superfields will in general have to satisfy differential constraints
in order to be made irreducible.

\subsection{Examples}

We will now illusrate this procedure with some examples.
Firstly we reconsider the hypermultiplet. This has dilation weight
$\D=1$, has no spin and no $R$-charge but transforms under the
fundamental of $SU(2)$ ie it has $a_1=1$. From these quantum numbers
one 
can calculate the Dynkin labels and put them
above the Dynkin diagram giving:
\be \begin{picture}(210,10)
\put(20,0){\makebox[0pt][l]{$\bt\hspace{3em}\ominus\hspace{3em}\bt
    \hspace{2.7em}
\ominus \hspace{3em}\bt$} \rule[.5ex]{15.em}{.1ex}

 }
\put(20,15){  0
\hspace{2.5em} 0 \hspace{2.5em} 1 \hspace{2.5em} 0 \hspace{2.5em}
0}
\end {picture}
\ee
This representation given as a superfield on various superspaces is
represented by the relevant Dynkin diagrams:
\be
\ba{ccc}
 \begin{picture}(130,10)(0,0)
 \put(20,0){\makebox[0pt][l]{$\bt\hspace{1em}\otimes 
    \hspace{1em}\bt
    \hspace{1em}
\otimes \hspace{.7em}\bt$} \rule[.5ex]{7.em}{.1ex}
 }
\put(17,10){  0
\hspace{.6em} 0 \hspace{.6em} 1 \hspace{.6em} 0 \hspace{.6em}
0}
\end {picture}
&
 \begin{picture}(130,10)(0,0)
\put(20,0){\makebox[0pt][l]{$\bt\hspace{1em}\otimes\hspace{1em}\times
    \hspace{.8em}
\otimes \hspace{.8em}\bt$} \rule[.5ex]{7em}{.1ex}
 }
\put(20,10){  0
\hspace{.6em} 0 \hspace{.6em} 1 \hspace{.6em} 0 \hspace{.6em}
0}
\end {picture}
&
 \begin{picture}(130,10)(0,0)
\put(20,0){\makebox[0pt][l]{$\bt\hspace{1em}\ominus\hspace{1em}\times
    \hspace{.7em}
\ominus \hspace{.7em}\bt$} \rule[.5ex]{7em}{.1ex}

 }
\put(20,10){  0
\hspace{.6em} 0 \hspace{.6em} 1 \hspace{.6em} 0 \hspace{.6em}
0}
\end {picture}
\\
\rm{Minkowski}&\rm{Harmonic}&\rm{Analytic}
\ea
\ee
We can read off from these Dynkin diagrams the transformation
properties  
of the hypermultiplet as a superfield on Minkowski, harmonic and
analytic superspace.

In particular the `1' above the central node
indicates that the hypermultiplet transforms under the fundamental
representation of
the internal group $SL(2)$ on Minkowski superspace. 
On harmonic superspace and analytic superspace on the other hand since 
there is a cross through the central node the hypermultiplet carries
no indices but has a non-trivial internal $\com$-charge. 

We know from section~\ref{Y} that the hypermultiplet satisfies
constraints as a superfield on Minkowski and harmonic superspace,
whereas it is unconstrained on analytic superspace due to analyticity
in the internal coordinates. This turns out to
be generally true: all unitary irreducible representations of the
superconformal group are given as superfields on
analytic superspace which are analytic but otherwise unconstrained. We
illustrate this with another example.

\subsection{Superfields with superindices}\la{transf}

As already hinted at, for more general representations the superfield may
transform linearly under a {{supergroup}}. 
One simply reads off the representation of the Levi subalgebra (under
which the fields transform linearly) from the 
super Dynkin diagram. One can then express this representation as a
tensor superfield by finding the corresponding (super) Young tableau
(according to a straightforward formula.) The number of boxes of the
Young tableau then indicates the number of superindices one needs and
also the symmetrisation of these indices.

We will illustrate this with another  simple example.
The $N=2$ super Maxwell multiplet can be described as a chiral
superfield on $N=2$ Minkowski superspace which also satisfies a
second-order constraint. However, it can also be given on analytic superspace
as follows. On calculating the Dynkin labels from the quantum numbers,
one obtains the following Dynkin diagram which gives the 
 Levi subalgebra indicated below it: 
\be \begin{picture}(210,90)(0,-70)
\put(20,0){\makebox[0pt][l]{$\bt\hspace{3em}\ominus\hspace{3em}\times
    \hspace{2.7em}
\ominus \hspace{3em}\bt$} \rule[.5ex]{15.5em}{.1ex}

 }
\put(20,15){  0
\hspace{2.5em} 1 \hspace{2.5em} 0 \hspace{2.5em} 0 \hspace{2.5em}
0}
\put(110,-20){$\Downarrow$}
\put(-10,-55){$\gl=$}
\put(20,-55){\makebox[0pt][l]{$\bt\hspace{3em}\ominus\hspace{3em}\times
    \hspace{2.7em}
\ominus \hspace{3em}\bt$} \rule[.5ex]{4.em}{.1ex}
 }
\put(155.5,-55){\rule[.5ex]{4.3em}{.1ex}}
\put(20,-40){  0
\hspace{2.5em} 1 \hspace{2.5em} 0 \hspace{2.5em} 0 \hspace{2.5em}
0}
\put(30,-75){$ \gs \gl(2|1) \hspace{50pt} \com \hspace{50pt} \gs\gl(2|1)$}
\end {picture}
\ee
The representations of the two $SL(2|1)$ supergroups are carried by
the superindices $A,A'$ respectively. We see from the lower Dynkin
diagram that although the analytic superfield carries a trivial
representation of the right $\gs \gl(2|1)$ (as indicated by the
zero Dynkin label) it carries a non-trivial representation of the left $\gs
\gl(2|1)$ (as indicated by the non-zero Dynkin label.) In fact the
diagram indicates that the superfield carries the anti-fundamental
representation of the left $\gs \gl(2|1)$.
This corresponds to a superfield $W_{A}=(W_{\a},W)$ with one
superindex.

The prescription of how to move from one patch of the Riemann sphere
to another patch is completely determined. For $y\rightarrow y'=1/y$, $W_A(x,\th^2,\vf_1,y) \rightarrow W'_A(x',\th'^2,\vf'_1,y')$
where
\be
\left( \ba{c} W'_\a\\W' \ea \right)=\left( \ba{c}
 W_{\a}/y\\-W_{\b}(\th^{2})^{\b}/y +W \ea \right),\quad
x'=x-{\th^2 \vf_1 \over y}, \quad \th'^2=-{\th^2\over y},\quad
\vf'_1={\vf_1\over y}.
\ee
Demanding holomorphicity in $y$ on both patches of $S^2$ then gives
the correct on-shell
components $(\r_{1\a},\r_{2\a}, F_{\a\b},C)$ all satisfying their
equations of motion:
\be \ba{rcl}
 W_{\a}&=&\r_{1\a}+y \r_{2\a}+(\th^2)^{\b}F_{\a\b} -
 (\vf_1)^{\adt}\del_{\a\adt}C\\&&-(\th^2)^{\b}(\vf_1)^{\bdt}\del_{\b\bdt}\r_{2\a}\\ 
 W&=&C-(\th^2)^{\a}\r_{2\a}
 \ea
\ee
Note that as in the case of the hypermultiplet, no differential
constraints are needed in order for the analytic superfield to carry
an irreducible representation.

So to summarise, in this section we have seen how one can give any superconformal
representation as a superfield on any
superspace (except super twistor spaces for which the construction
would lead to an infinite dimensional representation of the Levi
subgroup). 
In general the superfields require additional differential constraints 
in order to carry irreducible representations. However, {\em {all}} unitary irreducible representations (UIRs) can be given as
analytic superfields on analytic
superspace, which are otherwise
unconstrained~\cite{Heslop:2001zm,Heslop:2000mr}.

\section{$N=4$ SYM on $N=4$ analytic superspace}\la{sec:SYM}

From now on we specialise to $N=4$ analytic superspace which has
Dynkin diagram:
\be \begin{picture}(220,10)
\put(0,0){\makebox[0pt][l]{ $\bt\hspace{2em}\ominus
    \hspace{2em}\bt \hspace{2em} \times \hspace{2em} \bt 
    \hspace{2em}
\ominus \hspace{2em}\bt$}} 
\put(5,0){\rule[.5ex]{205pt}{.1ex}}
\end {picture}
\ee
It is a coset space of the $N=4$ superconformal group, $P\bsh
  SL(4|4)$, with coset representative $s(X)$
  where 
\be
 P={\tiny \left( \ba{cc|cccc|cc}
                    \hspace{-5pt}  \bt \hspace{-5pt}  &  \bt  &\bt \hspace{-5pt}&\bt \hspace{-5pt}&&&&\\
                      \hspace{-5pt}   \bt \hspace{-5pt}  &  \bt   &\bt \hspace{-5pt}&\bt \hspace{-5pt}&&&&\\
\hline                \hspace{-5pt}   \bt \hspace{-5pt}  &  \bt   & \bt \hspace{-5pt}& \bt \hspace{-5pt} &   & &&\\
                      \hspace{-5pt}  \bt \hspace{-5pt} &\bt   &\bt \hspace{-5pt} &\bt \hspace{-5pt} &&& &\\
                      \hspace{-5pt}     \circ \hspace{-5pt}  & \circ   & \circ \hspace{-5pt} & \circ \hspace{-5pt}   & \bt \hspace{-5pt}  & \bt 
                      &\hspace{-5pt}\bt \hspace{-5pt} &\hspace{-5pt} \bt \hspace{-5pt}
\\
 \hspace{-5pt} \circ \hspace{-5pt}  & \circ   & \circ \hspace{-5pt} & \circ \hspace{-5pt}   & \bt \hspace{-5pt}  & \bt   &\hspace{-5pt}\bt \hspace{-5pt} &\hspace{-5pt} \bt \hspace{-5pt}\\
\hline  \hspace{-5pt}\circ \hspace{-5pt}  & \circ  & \circ \hspace{-5pt} & \circ \hspace{-5pt}   & \bt \hspace{-5pt}  & \bt   &\hspace{-5pt}\bt \hspace{-5pt} &\hspace{-5pt} \bt \hspace{-5pt}\\
 \hspace{-5pt} \circ \hspace{-5pt}  & \circ   & \circ \hspace{-5pt} & \circ \hspace{-5pt}   & \bt \hspace{-5pt}  & \bt   &\hspace{-5pt}\bt \hspace{-5pt} & \hspace{-5pt}\bt \hspace{-5pt}\\
                        \ea \right)}
\qquad
s(X)= \left(\ba{cc}
1_{2|2}&X\\ 0_{2|2}&1_{2|2} \ea \right)
\qquad
 X^{AA'}=\left( \ba{c|c} x^{\a \adt} & \l^{\a a'} \\ \hline
                        \pi^{a\adt}  &  y^{a a'}  \ea
 \right).
\ee
As previously, here the nodes correspond to non-zero elements of $P$
and black nodes correspond to non-zero elements of $L$ under which
superfields transform.

Superfields on analytic superspace transform linearly under the Levi
subalgebra $\gl=sl(2|2) \oplus sl(2|2) \oplus \com$.

\subsection{$N=4$ SYM on analytic superspace}

The component fields of $N=4$ SYM, lie in a single  analytic superfield:
\be
(\f_{ij},\psi^i,F_{\m \n}) \rightarrow
W(x,y,\l,\p)
\ee
All operators in the {{free}} theory can be obtained on analytic
superspace by multiplying $W$'s, 
applying (super)derivatives and taking the trace over the gauge group $SU(N_c)$.
The simplest examples are the `chiral primary operators' or `CPOs',
given as
 \be A_p=Tr(W^p)\ee
which are also known as `half BPS' and are dual to Kaluza Klein states
on $S^5$ via the AdS/CFT correspondence.
\be
\begin{picture}(210,10)
\put(0,0){\makebox[0pt][l]{ $\bt\hspace{2em}\ominus
    \hspace{2em}\bt \hspace{2em} \times \hspace{2em} \bt 
    \hspace{2em}
\ominus \hspace{2em}\bt$}} 
\put(5,0){\rule[.5ex]{205pt}{.1ex}}
\put(0,10){0 \hspace{2em} 0 \hspace{2em} 0 \hspace{1.5em} p\hspace{2em}
    0 \hspace{2em} 0 \hspace{1.5em} 0}
\end {picture}
\ee
The simplest CPO, $T:=A_2$ contains the entire energy momentum
multiplet.

\subsection{Examples with superindices/ protected operators}

A more complicated example of an operator on $N=4$ analytic superspace 
is the Konishi operator. This is written in the free theory as 
\be \cK_{AB,A'B'}=\tr\left(\del_{(A A'}W\del_{B)B'}W+...\right)\ee
(the dots denote further total derivative terms needed to ensure the
operator transforms correctly). This operator is known to develop an
anomalous dimension in the interacting theory.

A seemingly similar operator is
\be\cO_{AB,A'B'}=\del_{(A A'}T\del_{B)B'}T+...
\ee
This operator is protected however from renormalisation, unlike the
Konishi operator. The question arises as to why these two operators
have such different properties. Analytic superspace provides a simple
way to answer this question and to find all such protected
operators~\cite{Heslop:2001dr}.
 {Both} $\cO$ and $\cK$ are {short} supermultiplets in the {free
   theory} (by short we
 simply mean that the operator does not have a full theta expansion
 when written as a superfield on $N=4$ Minkowski superspace).
However, $\cK$ cannot be extended to the interacting theory
  on analytic superspace (since 
  there is no covariant superderivative there\footnote{It can however
    be written abstractly on analytic superspace in terms of
    ``quasi-tensors''~\cite{Heslop:2001gp}}) whereas 
 $\cO$ {can} be extended to the interacting case (since $T$,
  being gauge invariant, only requires the normal derivative
  $\del$). Therefore $\cO$ must
  remain short in the interacting theory (since all tensor
  superfields on analytic superspace are irreducible and we know from
  the free theory that it must be short.)
Superconformal representation theory tells us that operators
  with anomalous dimensions must be long and so we conclude that 
$\cO$ cannot develop an anomalous dimension and hence must be
protected.

This then allows a very straightforward generalisation: all operators
written in terms of CPOs which are short in the free theory are
protected. 
So we see that analytic superspace  gives a straight forward way of
classifying protected operators.

In fact, one can also prove the non-renormalisation of protected
operators using correlation
functions~\cite{Arutyunov:2001qw,Heslop:2001gp} and this latter method
is the only way to prove non-renormalisation of
operators in the six dimensional (2,0) superconformal field theory
since there is no known classical theory one can use in the latter
case (see section~\ref{6d}).

\subsection{Correlation functions}

Using analytic superspace one can also completely solve the
superconformal Ward identities for
{all} correlation 
functions of gauge invariant operators.
{This is done by
adapting the Minkowski space techniques of~\cite{Osborn:1994cr}.}

The correlation functions are written in terms of the analytic
coordinates at points 1~to~$n$, $X_1, X_2 ,..X_n$.
The propogators are given as $g_{ij}=\rm{sdet} (X^{-1}_{ij})$.

The general solution of the Ward identities is then  given in the
following schematic form 
\be <12..n> =\Pi_{j=2}^n \cR_j(X_{1j}^{-1})\cR'_j(X_{1j}^{-1})\sum_t 
t^{\cR_2\ldots \cR_n;\cR'_2\ldots \cR'_n}_{\cR_1\cR'_1} P_t F_t\la{swi} \ee
where
$<12..n> :=<\cO_{\cR_1 \cR'_1}(X_1) ...>$,
 $\cR_i$ are $SL(2|2)$
  representations (specified using superindices),
 $t$ is a tensor which is a monomial in $X_{12k}=
X_{12}X^{-1}_{2k} X_{k1}$ and  their inverses, 
$F_t$ is an arbitrary invariant which depends on the coupling
  constant and $P_t$ are monomials of the propogators $g_{ij}$. Note
  that there are no non-trivial invariants for $n\leq 3$.
The problem of finding the invariants has also been completely solved~\cite{Heslop:2003xu}.

One has to check that the resulting correlator is analytic  in
  the internal coordinates: if it is not then that correlator must be
  ruled out.

We illustrate the formula~\eq{swi} with a couple of examples.
The simplest cases are the correlators of CPOs, given by a monomial of
the propogators, $P$,
times an invariant: $<A_{p_1} A_{p_2} .. A_{p_n}> =P
F$~\cite{Howe:1995aq,Howe:1996rm}.

The simplest example with superindices is 
$<\cO_{AA'} T T>=P \ t_{AA'}$
where $t$ is uniquely given by the monomial $t^{AA'}=c
(X^{-1}_{123})_{AA'}$ where c is a constant.     
  
Using formula~\eq{swi} together with the reduction formula and
properties 
of correlators under an additional $U(1)_Y$ symmetry~\cite{Intriligator:1998ig}, it can be proven
that all two- and three-point functions 
of all protected operators are independent of the coupling 
constant~\cite{Heslop:2001gp}.

\section{From analytic superspace to harmonic superspace} \la{sec:hss}

We here show how one can lift an analytic superfield (with superindices) to a
harmonic (in general,, constrained) superfield. 

The harmonic superspace is a fibration over analytic superspace. This
fibration splits into two parts corresponding to the two supergroups
$SL(2|2)$ under which the analytic superfields transform. One thus
obtains left and right coset spaces with representatives
\be s(\r)^A{}_B=\left( \ba{c|c} \d^{\a}{}_{\b} & \r^{\a}{}_b\\
\hline 0&\d^a{}_b\ea \right) \qquad s'(\h)_{A'}{}^{B'}=\left(
\ba{c|c} \d_{\a'}{}^{\b'}& 0\\ \hline
\h_{\a'}{}^{b'}&\d_{a'}{}^{b'}\ea \right). 
\ee
Here $\r$ and $\h$ will become the extra odd coordinates that harmonic
superspace has above analytic superspace. 
To
obtain a superfield on harmonic superspace from one on analytic
superspace, simply multiply the analytic superfield on the right
by $\cR(s^{-1}(\r))$ and on the left by $\cR'(s'(\h))$ (where $\cR$
and $\cR'$ are the representations of the two $GL(2|2)$ supergroups
under which the analytic superfields transform) and choose the component
with the 
maximum number of internal indices. This is best shown with some examples.

A one-half BPS state $A$ has no indices, so it lifts
trivially to harmonic superspace. It does not depend on the extra
$(\r,\h)$ coordinates as we would expect.
A one-quarter BPS operator with Dynkin labels of the
form $[001d100]$  is given on analytic superspace
by a superfield, $V_{A'A}(x,\l,\p,y)$ with two superindices.  It lifts
to a
superfield $v_{a'a}(x,\l,\p,y,\r,\h)$ in harmonic superspace where
\be
\ba{rcl}v_{a'a} &=& s'_{a'}{}^{B'}V_{B'B}(s^{-1})^B{}_a\nn\\
 & =&V_{a'a} -V_{a' \b} \r^{\b}{}_a + \h_{a'}{}^{\b'}V_{\b' a} -
\h_{a'}{}^{\b'}V_{\b' \b}\r^{\b}{}_a .\ea
\ee
The constraints on $v_{a'a}$ then follow straightforwardly from the
dependence on the `extra' coordinates $\p,\eta$. In this way one can
see that from this point of view the origin of the constraints of
infinite dimensional superconformal representations on
harmonic superspace (and hence also on Minkowski superspace) comes
simply from the constraints of finite dimensional representations of
$GL(2|2)$.

\section{Analytic superspace in six dimensions}\la{6d}

One can also consider superconformal field theories in six dimensions
from an analytic superspace point of view.

The (complexified) $d=6$ $(N,0)$ superconformal group is $Osp(8|2N)$
which has maximal bosonic subgroup $SO(8) \times Sp(2N)$.
The corresponding Lie algebra $o s p (8|2N)$ is simply the set of
$(8|2N)\xz (8|2N)$ supermatrices $M$ s.t. 
\be
osp (8|2N) =\{M | M J + JM^{ST} = 0 \} \qquad \qquad
 J= {\tiny \left( \ba{cc|cc} 0_4\hspace{-5pt} &1_4 & 0\hspace{-10pt}& 0 \\
                            1_4 \hspace{-5pt}&0_4 & 0\hspace{-10pt}& 0 \\
\hline                       0 \hspace{-5pt}& 0 &0_N \hspace{-10pt}& -1_N \\
                             0\hspace{-5pt}& 0  &1_N\hspace{-10pt}& 0_N \ea \right)}
\ee
where $M^{ST}$ denotes the supertranspose.
The corresponding Dynkin diagram is
\be
\begin{picture}(320,20)(0,20)
\unitlength=2pt
\put(10,10){$\bt$}\put(30,10){$\bt$}\put(50,10){$\bt$}
\put(70,10){$\circ$}\put(90,10){$\bt$}\put(110,10){$\bt$}
\put(116,10){$\cdots$}\put(130,10){$\bt$}\put(150,10){$\bt$}
\put(12,11.5){\line(1,0){58.5}}
\put(72.5,11.5){\line(1,0){38}}
\put(131,11){\line(1,0){20}}\put(131,12){\line(1,0){20}}
\put(93,13){$\overbrace{\phantom{a \hspace{100pt} a}}^N$}
\end{picture}
\ee
As in the 4d case crosses can be placed arbitrarily on the Dynkin diagram to
  represent superspaces. 





\subsection{$d=6 \ (2,0)$ superconformal symmetry }

We would like to apply these techniques in particular to the $(2,0)$
superconformal field theory first considered in~\cite{Howe:1983fr} and
reformulated into harmonic superspace in~\cite{Howe:1998jw}.
This is a somewhat mysterious theory which is dual to M theory on
$AdS_7 \times S^4$ via the AdS/CFT correspondence. 
Superconformal symmetry provides a possible method to study properties of this theory. 

We will use the simplest analytic superspace which has Dynkin
diagram. 
\be
\begin{picture}(240,10)(0,20)
\unitlength=2pt
\put(10,10){$\bt$}\put(30,10){$\bt$}\put(50,10){$\bt$}
\put(70,10){$\circ$}\put(90,10){$\bt$}
\put(107,8.5){\Huge$\xz$}
\put(110,10){$\bt$}
\put(11.5,11.5){\line(1,0){59}}
\put(72.5,11.5){\line(1,0){19}}
\put(92,12){\line(1,0){20}}
\put(92,11){\line(1,0){20}}
\end{picture}
\ee
We can read off the parabolic subalgebra $\gp$ and a coset representative
$s$
\be
\gp=\left\{ \left( \ba{cc} -A^A{}_B &0 \\
                         - C_{AB}   &D_{A}{}^B \ea \right) \right\}
\qquad \qquad
s(X)=\left( \ba{cc} 1 &X\\
                    0 &-1 \ea \right)
\quad X=\left( \ba{c|c} x^{\a \b} & \l^{\a b} \\
       \hline         -\l^{a \b}& y^{ab} \ea \right)
\ee
where 
$D_{A}{}^{B}= (-1)^{A(A+B)} A^{B}{}_A$ and
 $A=(\a,a)$ is a superindex, $\a=(1,2,3,4)$ $a=(1,2)$.

One can then proceed largely by analogy with the four dimensional
case. The simplest superfield on analytic superspace is a scalar with
charge 1 and is denoted $W$. This contains all the fundamental fields
of the theory. Given that there is no known classical theory
one can only form composite operators explicitly in the free theory.
However, one can still abstractly consider superfields as
representations of the superconformal group.

It is possible to prove that certain operators which lie on the threshold
of the unitary bounds and which lie in the OPE of two short operators
can not have anomalous
dimensions, by considering restrictions on the three-point functions of
these operators and the two short operators. 

One can form correlation functions and find their Ward identities which
can then be solved analagously to the four dimensional case shown above.   
It is particularly interesting to study the 4-point function $<TTTT>$ and
compare with the analogous correlator in $N=4$ SYM.
Arutyunov and Sokatchev have shown that this can be written in terms
of a single function of two variables~\cite{Arutyunov:2002ff} in
contrast to the four 
dimensional case where one also needs a 
(non-renormalised) function of 1 variable~\cite{Eden:2000bk}. Using analytic
superspace one can write down the complete four-point function of all
operators in the energy momentum multiplet in a compact formula, one
finds that the solution of the superconformal Ward identities can be
solved in terms of a single function of two-variables even before
crossing symmetry is taken into account. However, the relation of this
function to operators appearing in the OPE of two Ts via the conformal
wave expansion is more complicated than in the four dimensional case~\cite{Heslop:prog}. 


\ed